%% file: paper.tex
\documentclass[runningheads]{llncs}

\usepackage[T1]{fontenc}
\usepackage{graphicx}
\usepackage[dvipsnames]{xcolor}
\usepackage{tikz}
\usepackage{pgfplots}
\usepackage{pgfplotstable}
\usepackage{booktabs}
\usepackage{subcaption}
\usepackage{amsmath,amsfonts}
\usepackage{amssymb}
\usepackage{mathrsfs}
\usepackage{contour}
\contourlength{1.4pt}
\usepackage[final,expansion=false]{microtype}
\usepackage{hyperref}

\pgfplotsset{compat=1.18}
\usetikzlibrary{calc, shapes, patterns, arrows.meta, backgrounds}

\definecolor{schwellwert}{RGB}{0,114,178}
\definecolor{gitter}     {RGB}{213,94,0}
\definecolor{gewichtet}  {RGB}{0,158,115}
\definecolor{komitee}    {RGB}{204,121,167}

\definecolor{cnormal}{RGB}{0,114,178}
\definecolor{ctentative}{RGB}{230,159,0}
\definecolor{creadonly}{RGB}{204,121,167}

\definecolor{fab}{HTML}{bebada}
\newcommand{\para}[1]{\smallskip\noindent\textbf{#1}}

\usepackage{todonotes}
\presetkeys{todonotes}{
  inline,
  color=yellow!20
}{}
\setuptodonotes{inline}

\usepackage{multirow}
\usepackage{makecell}

\usepackage{tcolorbox}
\newcommand{\acceptancenote}{%
  Accepted at LADC 2026, the 15th Latin-American Symposium on Dependable and Secure Computing (\url{https://ladc.sbc.org.br/2026/}).
  This is the author's version (submitted version) of the paper.
  DOI: to be inserted after publication.%
}
\AddToHookNext{shipout/foreground}{%
  \put(\dimexpr 1in+\hoffset+\oddsidemargin\relax,
       -\dimexpr 1in+\voffset+\topmargin+\headheight+\headsep+\textheight+20pt\relax){%
    \makebox(0,0)[lt]{%
      \begin{minipage}[t]{\textwidth}
        \begin{tcolorbox}[colback=black!5, colframe=black!20, boxrule=0.4pt, arc=1mm,
                          left=5pt, right=5pt, top=3pt, bottom=3pt, before skip=0pt, after skip=0pt]
          \normalfont\footnotesize\acceptancenote
        \end{tcolorbox}
      \end{minipage}}}}

\begin{document}
\title{Exploring Adaptive Byzantine Quorum Systems to Improve Latency in the WAN}

\titlerunning{Exploring Adaptive Byzantine Quorum Systems in the WAN}

\author{Linus Gnan \and
        R\"udiger Kapitza \and
        Christian Berger}
\authorrunning{L. Gnan et al.}
\institute{Friedrich-Alexander-Universit\"at Erlangen-N\"urnberg, Germany\\
\email{\{linus.gnan,ruediger.kapitza,christian.g.berger\}@fau.de}}

\maketitle

\begin{abstract}
Quorum systems enforce strict consistency in Byzantine fault-tolerant (BFT) state machine replication: Before a value is decided, a subset of replicas (called \textit{quorum}) must exchange votes for the value. In wide-area networks, the size and composition of a quorum determines the speed at which replicas can make progress and thus impacts the latency perceived by clients. A variety of quorum constructions has been proposed, e.g., threshold, weighted, grid and others, but the literature offers little in the way of an apples-to-apples comparison. Each construction typically lives inside a different protocol implementation, so the influence of quorum design on client latency cannot be easily separated from other implementation effects, especially the use of optimization techniques. In this paper, we close that gap by extending BFT-SMaRt with a modular quorum-system abstraction that allows a variety of quorum constructions to live inside the same framework and interact with the same set of optimization techniques. By using our framework, we explore the impact of different quorum system constructions on client-observed latency in wide-area BFT replication.

\end{abstract}


\section{Introduction}

Distributed services may need to survive Byzantine failures of individual machines. A common approach to build fault-tolerant systems on top of a set of independently failing server replicas is the state machine replication (SMR) approach~\cite{schneider1990implementing}. Modern BFT SMR libraries, such as BFT-SMaRt~\cite{bessani2014state}, make it easy to develop fault-tolerant services, and can, for example, be used as \textit{ordering service} in permissioned blockchain platforms like Hyperledger Fabric~\cite{sousa2018byzantine} where replicas may be deployed in a globally distributed infrastructure.

The core of a Byzantine state machine replication protocol is often based around a total-order broadcast, for which a \textit{Byzantine dissemination quorum system} is essential~\cite{malkhi1998byzantine}: It defines all possible \textit{quorums} (sets of replicas), such that two quorums will always have an intersection that is not entirely composed of Byzantine replicas and at least one quorum will always be available to access despite any failure scenario happening. While the intersection property of quorum systems ensures consistency, the availability property is necessary to allow the BFT protocol to progress even when failures occur.

\begin{figure}[t]
  \centering
  \begin{subfigure}{0.54\linewidth}
    \centering
    \includegraphics[width=1\linewidth]{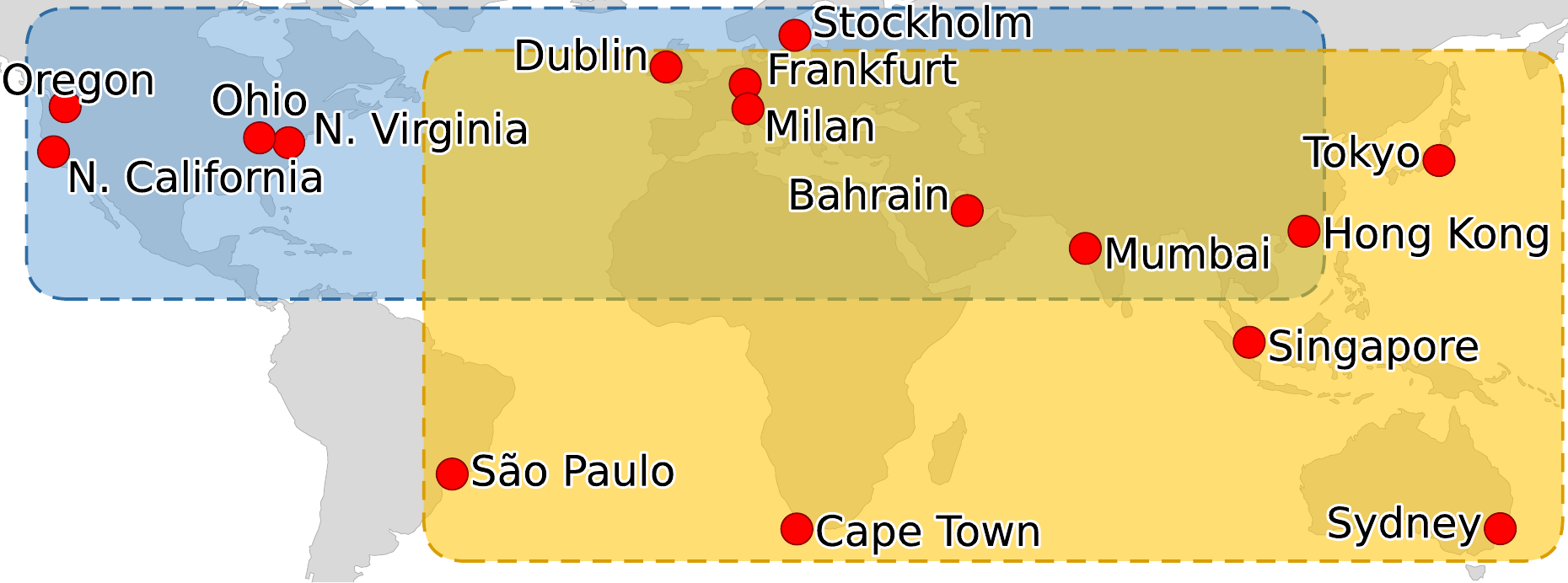}
    \caption{Geo-replicated system deployment.}
    \label{fig:deployment}
  \end{subfigure}\hfill
  \begin{subfigure}{0.46\linewidth}
    \centering
    \resizebox{\linewidth}{!}{%
    \input{figures/normalbetrieb.tex}
    }
    \caption{BFT-SMaRt message pattern.}
    \label{fig:BFT-consensus}
  \end{subfigure}
  \caption{(a) Two Byzantine dissemination quorums (see \colorbox{SkyBlue}{blue}  and \colorbox{yellow}{yellow}) in a WAN intersect in at least one correct replica (here:~$n=16,f=5$). These quorums are used in BFT-SMaRt's message pattern: (b) the votes of the \colorbox{SkyBlue}{blue} quorum in the \textsc{Write} and \textsc{Accept} stages are sufficient to make progress. Depending on the optimizations used, quorums are also necessary at the client side.}
  \label{fig:BFT}
\end{figure}

\para{Quorum systems.}
Quorum systems can be defined by quorum formation rules. The most widely employed rule in BFT systems~\cite{castro1999practical,bessani2014state,hotstuff19,danezis_narwhal_2022,babel_mysticeti_2025} is one that states that \textit{any subset of $2f+1$ out of $3f+1$ replicas forms a quorum}, which is known as the threshold quorum system. A replica advances to the next protocol stage depending on the slowest of messages arriving from others necessary to complete its quorum. In a wide-area (WAN) deployment, this means that the fastest group of replicas that form a quorum can determine the speed at which the protocol makes progress. Thus, the choice of quorum system  (and the resulting quorums) impacts consensus latency and is also directly felt by clients. For instance, the diameter (i.e., the maximum pairwise replica-replica latency) of the blue quorum in Fig.~\ref{fig:BFT} is 132\,ms while it is 205\,ms for the yellow quorum.

Several quorum constructions were studied in isolation. The classic \textit{threshold} quorums used in protocols like PBFT~\cite{castro1999practical}, BFT-SMaRt~\cite{bessani2014state} and HotStuff~\cite{hotstuff19} require a threshold, typically more than $2/3$ of all replicas. \textit{Weighted} quorums as shown by WHEAT~\cite{sousa2015separating} can be used to form proportionally smaller quorums by assigning higher voting weight to a group of well-connected replicas. This idea can accelerate quorum formation and thus speed up consensus latency. \textit{Grid} quorums~\cite{malkhi1998byzantine} arrange replicas in a square grid and intersect through full rows and columns. Fast Byzantine Paxos (FaB)~\cite{martin2006fast} reduces the number of communication steps for agreement by one step at the cost of decreased resilience \textit{and} a proportionally larger threshold quorum system ($4f+1$ out of $n=5f+1$). Interestingly, we show that the side effect of having proportionally larger quorums can offset the latency gains achieved by saving a  communication step. \textit{Adaptive quorum systems} are quorum systems that can be automatically  tuned during runtime, e.g., finding the concrete weight assignment that optimizes consensus latency~\cite{berger20aware}.

What we aim for in this paper, however, is an extensive \textit{apples-to-apples} comparison.
Typically, a quorum system is evaluated inside some specific BFT protocol implementation, with its own
set of optimizations and replica-client behavior. Differences in client-observed latency between individual  quorum systems risk being attributed to the quorum system when they are in fact rooted in message patterns, implementation-specific optimizations or geography.

\para{Contributions.} We close the gap by extending BFT-SMaRt
with a modular quorum-system abstraction and re-implement four quorum constructions on top of it, namely threshold, weighted, grid and committee quorums. All share the same consensus core and the same \emph{read-only} and \emph{tentative} optimizations~\cite{castro1999practical,bessani2014state}, and the same runtime self-optimization mechanisms inherited from AWARE~\cite{berger20aware}. As a fifth point of comparison we implement the message pattern of the FaB Paxos consensus protocol with its larger threshold quorum system, which allows us to study the trade-off between larger quorum size and fewer communication steps. We evaluate this framework using a global AWS deployment scenario of $n=16$ replicas, spanning all continents (see Fig.~\ref{fig:deployment}), using a statistically recorded latency matrix and a deterministic network simulation~\cite{jansen2022co} for improved reproducibility. Moreover, we like to mention four main observations upfront:

\begin{itemize}
    \item \textit{One step less is not always faster}. The FaB Paxos protocol uses a message pattern that shortens consensus by one step but expects $n\geq 5f+1$ replicas and a proportionally larger quorum    ($\lceil\frac{n+3f+1}{2}\rceil$ instead of $\lceil\frac{n+f+1}{2}\rceil$). In a WAN, the wait for the larger quorum can offset the latency gain from the saved communication step and lead to a \textit{higher client-observed latency} in several regions than that of the normal 3-step protocol, even when the \textit{consensus latency} observed by the leader appears to be shorter.

\smallskip
    \item The \textit{weighted} and \textit{committee} quorum systems make it possible to efficiently trade resilience for lower latency by forming proportionally smaller quorums and achieving the best results overall with a slight edge for committee-based quorums. This effect is not observed for threshold quorums, whose size also depends on $f$ but can never shrink in size below $50\%$ of replicas. Grid quorums do not pay off in the WAN setting and lead to the highest client latency.

\smallskip
    \item \textit{The right consensus-side quorum is not always the right client-side quorum.}
    Once linearizability under \emph{read-only}  or \emph{tentative} optimizations forces clients to collect a full quorum (instead of a weak certificate of $f+1$ responses), the geographic distribution of clients becomes very important.
    Not all client regions might equally benefit from a fast-progressing consensus quorum.

\smallskip
    \item \textit{Abstracting the quorum system correctly is challenging}. In a feature-rich BFT SMR framework, abstracting the quorum system and allowing it to change during runtime can have many  (potentially breaking) side-effects on different protocol modules (e.g., leader-change and state transfer). We will describe how this challenge can be tackled without violating  correctness.
\end{itemize}

\section{Background}
\label{sec:bg}

To lay the groundwork for the remainder of this paper, we first review background on BFT frameworks, their optimizations as well as quorum systems.

\subsection{BFT Protocols}
\label{sec:bg:bft}
\para{PBFT \& BFT-SMaRt.} Castro and Liskov introduced PBFT~\cite{castro1999practical}, the first practical
BFT protocol. It
assumes $n=3f+1$ replicas to tolerate $f$ Byzantine faults,
needs partial synchrony~\cite{dwork1988consensus} for liveness, stays consistent under asynchrony
and achieves performance comparable to non-replicated systems
using a set of optimizations, e.g., batching, read-only requests, and tentative executions.
BFT-SMaRt~\cite{bessani2014state} is a multi-threaded, modular
replication library that implements the Mod-SMaRt protocol~\cite{sousa2012byzantine}
which layers a Byzantine consensus primitive (the protocol proposed by Cachin~\cite{cachin2009yet})
under a total-order broadcast protocol.
A client first sends its request to all replicas, then the
 \textit{normal-case pattern} proceeds
 in 3 steps (see Fig.~\ref{fig:BFT}):  a \textsc{Propose} by the leader, followed by all-to-all \textsc{Write} and \textsc{Accept} phases in which each replica waits for a quorum of identical votes before advancing. Without the read-only optimization, a client accepts a result after a weak certificate of $f+1$ matching replies.  Otherwise it waits for a full quorum~\cite{castro_pbft_thesis_2001}  of $\lceil \frac{n+f+1}{2} \rceil$ replicas to ensure  \textit{linearizability}~\cite{herlihy1990linearizability}.

\para{WHEAT \& AWARE.}
The WHEAT~\cite{sousa2015separating} protocol extends BFT-SMaRt for WAN optimization, integrating
tentative executions~\cite{castro1999practical},
and a bimodal weighted quorum system (detailed later)
in which
proportionally smaller quorums
of well-connected replicas emerge which accelerates consensus.
AWARE~\cite{berger20aware} adds latency-driven self-optimization during runtime:
Replicas monitor point-to-point latencies,
agree on a common latency matrix through total-order broadcast, and
periodically run a deterministic prediction model
to search for the leader location and weight distribution that minimize consensus latency. For larger replica sets, AWARE uses simulated annealing~\cite{Kirkpatrick671} to tune the weight distribution and leader.

AWARE's search is limited to the leader location and the weight distribution of a weighted quorum system; it can neither configure another construction nor exchange the construction itself.

\para{BFT Optimizations.} Besides batching, two optimizations matter here.
The \emph{read-only} optimization allows clients to skip the total-order broadcast: Their read requests are directly executed and responded by all replicas. To preserve linearizability~\cite{herlihy1990linearizability}, however,
\emph{every} request (write and read) now requires a response quorum at the client~\cite{castro_pbft_thesis_2001} rather than a weak certificate of $f+1$ matching responses. In \emph{tentative executions}, replicas execute and deliver the result to the client immediately after the \textsc{Write} phase, with the \textsc{Accept} phase running asynchronously~\cite{castro1999practical}. The client must collect a quorum, and the tentative state must be rolled back in the case that the \textsc{Accept} phase fails and the leader changes.

\subsection{Dissemination Quorum Systems}
\label{sec:bg:dqs}
Malkhi and Reiter~\cite{malkhi1998byzantine} provide formal notions on quorum systems: Let $U$ be the set of $n$ replicas. A quorum system $\mathscr{Q}\subseteq 2^U$ is
a non-empty family of pairwise-intersecting subsets of $U$. A
fail-prone system $\mathscr{B}\subseteq 2^U$ collects the possible
sets of faulty replicas we consider. For Byzantine faults with up to $f$ failures,
$\mathscr{B} = \{B \subseteq U : |B|=f\}$. A quorum system $\mathscr{Q}$ is called a \textit{dissemination quorum system} for a fail-prone system $\mathscr{B}$ if the following two properties are satisfied~\cite{malkhi1997byzantine}:
	\begin{enumerate}
	 \item $\textbf{D-Consistency:~}  \forall Q_1, Q_2 \in \mathscr{Q}, \forall B \in \mathscr{B}: Q_1 \cap Q_2 \nsubseteq B $
	 \item $\textbf{D-Availability:~}  \forall B \in \mathscr{B}, \exists Q \in \mathscr{Q}: B \cap Q = \emptyset $
	\end{enumerate}

\noindent Consistency guarantees intersection in at least one \textit{correct} replica.
Availability guarantees
a quorum can be accessed even when replicas are faulty.
For this paper, the term \textit{quorum system} always refers to dissemination quorum systems.

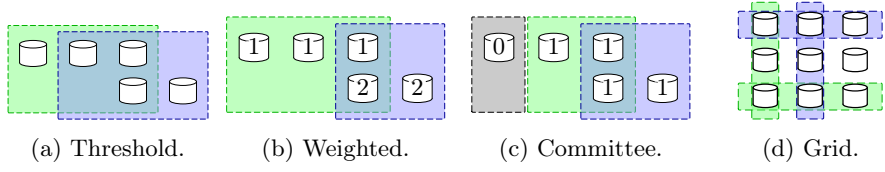
\begin{figure}[t]
  \centering
  \begin{subfigure}{0.22\linewidth}
    \centering
    \  \resizebox{\linewidth}{!}{%
    \input{figures/threshold-quorum}
    }
    \caption{Threshold.}
    \label{fig:quorum:threshold}
  \end{subfigure}\hfill
  \begin{subfigure}{0.24\linewidth}
    \centering
    \resizebox{\linewidth}{!}{%
    \input{figures/weighted-quorum}
    }
    \caption{Weighted.}
    \label{fig:quorum:weighted}
  \end{subfigure} \hfill
   \begin{subfigure}{0.24\linewidth}
    \centering
    \resizebox{\linewidth}{!}{%
    \input{figures/committee-quorum}
    }
    \caption{Committee.}
    \label{fig:quorum:committee}
  \end{subfigure}
   \begin{subfigure}{0.24\linewidth}
    \centering
    \resizebox{0.67\linewidth}{!}{%
    \input{figures/grid-quorums3x3}
    }
    \caption{Grid.}
    \label{fig:quorum:grid}
  \end{subfigure} \hfill
  \caption{Different quorum system constructions under study.}
  \label{fig:quorum-systems}
\end{figure}

\section{Quorum Constructions under Study}
\label{sec:bg:constructions}

In this section, we detail the quorum systems that we explore (see Fig.~\ref{fig:quorum-systems}).%

\para{Threshold quorums} are defined by a certain threshold $t$ in proportion to the system size $n$: For instance, any set of
$\lceil\tfrac{n+f+1}{2}\rceil$ replicas~\cite{malkhi1998byzantine} forms a quorum in BFT-SMaRt~\cite{bessani2014state}. Thus, quorum formation depends only on \emph{how many} replicas have voted/replied but not on \textit{which ones}. FaB~\cite{martin2006fast}
uses $n\geq 5f+1$ to save one communication step at the sacrifice of optimal resilience and it also requires collecting a proportionally larger quorum threshold of $\lceil\tfrac{n+3f+1}{2}\rceil$ replicas.

\para{Weighted quorums} as proposed by WHEAT~\cite{sousa2015separating} use $n=3f+1+\Delta$ replicas for some $\Delta>0$. While $2f$ replicas receive
$V_{max}=1+ \frac{\Delta}{f}$ weight, the remaining $f+\Delta+1$ replicas receive $V_{min}=1$. A quorum is any set of replicas whose accumulated weight reaches $2fV_{max}+1$. Using this scheme, the size of quorums varies from $2f+1$ (all
 $V_{max}$-replicas plus any single $V_{min}$-replica) to $n-f$ (the fallback if all faulty replicas have $V_{max}$). To make this quorum system \textit{adaptive}, AWARE~\cite{berger20aware} assigns the $V_{max}$ weights to the fastest clique of replicas during runtime, so that the small quorum is the one normally used to drive progress.

\para{Grid quorums} arrange $n=k^2$ replicas in a $k\times k$ grid with rows $R_1,\ldots,R_k$ and columns $C_1,\ldots,C_k$ to
construct a quorum from one column and $f+1$ rows~\cite{malkhi1998byzantine}:
\begin{equation} \mathscr{Q} = \{ C_j \cup \bigcup_{i\in I} R_i : I,{j} \subseteq {1...\sqrt{n}}, |I| = f+1 \} \end{equation}
\noindent This
is not minimal:
Since two such quorums intersect in at least two replicas per selected row,
 one column plus $r=\left\lceil\frac{f+1}{2}\right\rceil$ rows already satisfies D-Consistency.
Meanwhile, D-Availability requires that, despite $f$ faulty replicas, one fault-free column and at least $r$ fault-free rows remain. Thus, we require $r+f\leq k=\sqrt{n}$. By symmetry, it is also possible to allow constructions consisting of one row and $r$ columns. Thus, replicas may be assigned to grid positions adaptively, and the faster of the row- and column-oriented quorums can be used to make progress. Further, the assignment of replicas to grid positions can be chosen (meaning \textit{optimized}) during runtime. The resulting quorum system is presented in Eq.~\eqref{eq:grid-adaptive}.
\begin{equation}
\label{eq:grid-adaptive}
\mathscr{Q} = \{ C_j \cup \bigcup_{i\in I} R_i : I,{j} \subseteq {1...\sqrt{n}}, |I| = r \}
\cup \{ R_i \cup \bigcup_{j\in J} C_j : J,{i} \subseteq {1...\sqrt{n}}, |J| = r \}
\end{equation}

\para{Committee quorums} restrict quorum formation to a subset of well-connected replicas (called \textit{committee}). Replicas outside the committee participate as non-voting observers (i.e., they are not acceptors but only learners during consensus).
 For $n=3f+1+\Delta$, a committee $K\subseteq U$ contains $3f+1$ replicas. Members of $K$ receive voting weight $V_{max}=1$, whereas all other replicas receive weight $V_{min}=0$. A quorum requires accumulated weight $2f+1$ and thus consists of $2f+1$ committee members. Note that this weighting scheme yields the same minimal quorum size as in WHEAT, but allows a missing or failing committee member to be replaced by just a single other committee member (which is ideally located close to the group of other committee members). To optimize this quorum system during runtime, we let AWARE select the fastest $2f+1$ replicas to become part of the committee, then heuristically add the latency-wise closest replicas to the committee until the committee size reaches $3f+1$.

\section{A Modular Adaptive-Quorum-Systems Framework}
\label{sec:design}

Our framework aims to make the quorum system an exchangeable component of BFT-SMaRt. Doing so raises the questions that structure this section: How to \textit{encapsulate} a quorum system behind a single interface (Problem~1), how to let \textit{any} such quorum system be \textit{tuned} at runtime by AWARE (Problem~2), and how to integrate quorum-system \textit{transitions} correctly with respect to the rest of the protocol (Problem~3). Table~\ref{tab:bft-quorums} summarizes the resulting system variants after extending BFT-SMaRt and the quorums required for replicas and clients.

\para{Encapsulating the quorum logic (Problem~1).}
Independent of construction, the consensus protocol only ever asks a quorum system one question: \textit{Given these votes  collected so far, do their senders already form a quorum?}
So we model the quorum system as an abstract type (\textit{QuorumSystem}) which only answers this question and hides how a quorum is actually defined. In BFT-SMaRt's agreement protocol~\cite{sousa2012byzantine}, a replica waits for a quorum of identical votes in both \textsc{Write} and \textsc{Accept} phase (see Fig.~\ref{fig:BFT}). Whether a quorum is reached is evaluated once per incoming consensus message. Here, we replace the original  $\lceil \frac{n+f+1}{2} \rceil$ counting check with a call to our abstraction. A client also calls a special method  of \textit{QuorumSystem} when it collects responses. Depending on the optimizations used, this method either checks whether a weak certificate of $f+1$ matching replies was collected or whether a full quorum is available. The context that a quorum system needs ($n$, $f$, and any weights or grid layout) is moved into BFT-SMaRt's \textit{view}: A view has a \textit{QuorumSystem} object, which in turn includes the quorum system configuration. Since an updated view is always forwarded to clients once clients send a request with outdated view number, clients automatically obtain the current quorum system and evaluate quorums with the same logic as the replicas. Adding a quorum system (see Table~\ref{tab:bft-quorums}) then just means subclassing the abstraction and defining suitable methods for \textit{quorumFullFilled} and \textit{reconfigure}. The latter is necessary to recompute the quorums whenever $n$ or $f$ change.

\para{Tuning a quorum system at runtime (Problem~2).}
AWARE~\cite{berger20aware} optimizes two things using its measured latency matrix: First, the leader location (useful to every system variant in Table~\ref{tab:bft-quorums}). Second, the quorum system \textit{configuration} for quorum systems that can be configured. Examples of quorum configurations include the concrete weight distribution of WHEAT~\cite{sousa2015separating}, a specific grid layout, or defining the set of committee members.
\textit{Adaptive} quorum systems (like weighted, grid, committee) can adjust this configuration during runtime, which is integrated into AWARE: A \textit{candidate} configuration is evaluated by predicting its consensus latency through a simulated protocol run under the given network characteristics (the latency matrix) and the quorum formation rules defined by the chosen quorum system and used configuration. AWARE searches for the best candidate configuration. The search itself uses simulated annealing~\cite{Kirkpatrick671} to explore the search space for larger $n$. For this purpose a quorum system configuration type needs to implement two methods: a \textit{neighbor generator} (e.g., swap the weights of two replicas, or
swap two grid positions) which is driven by a seeded pseudo-random source, so that all correct replicas explore the identical sequence and deterministically agree on the same optimum; and a method that returns selectable leader candidates to shrink the search space. When a better configuration is found, each replica installs it inside a new view with an incremented view number. Note that this procedure is deterministic, because all replicas run the search at the same pre-defined consensus rounds using a consistent latency matrix (because latencies are agreed with total-order broadcast).

\begin{table}[t]
    \centering
    \resizebox{\textwidth}{!}{
        \input{tables/table1}
    }
    \vspace{1em}
    \small
    \textit{Legend: ro = read-only optimization, t = tentative execution,   col. = column.}
     \caption[Comparison of Different Quorum Variants]
    {Comparison of BFT-SMaRt system variants  with respect to their employed quorum system and enabled optimizations.
    }
    \label{tab:bft-quorums}
    \vspace{-0.35cm}
\end{table}

\para{Integrating quorum-system transitions correctly (Problem~3).}
Allowing the quorum system to change at runtime problematically interacts with two of BFT-SMaRt's sub-protocols that would implicitly assume a fixed notion of quorum: the \textit{leader change} and \textit{state transfer}. Each decided consensus instance has a \textit{proof}, which is a set of signed \textsc{Accept} messages whose senders form a quorum~\cite{sousa2012byzantine}. Verifying a proof means checking its signers \textit{are} a quorum.
During a BFT-SMaRt synchronisation phase (leader change), each replica reports its most recent decided consensus instance together with a corresponding proof, and the new leader selects the highest consensus instance with valid proof as the common, up-to-date synchronisation point. Only \textit{that} proof is included in the new-leader (\textsc{Sync}) message and re-validated by every replica before they install the new regency. A replica that lags behind this point does not obtain the proof from the leader change at all. Instead, it fetches the corresponding state via the \textit{state-transfer}  protocol, which likewise validates the proof attached to the transferred state.
The problem is now that an old proof (using a quorum $Q_{old}$) may have been produced under a quorum system $\mathscr{Q}_{old}$ that is no longer the current one $\mathscr{Q}_{new}$.
When such a proof ($Q_{old}$) is validated naively against the current quorum system, then a correct proof could be rejected ($Q_{old} \notin \mathscr{Q}_{new}$) and the most recent consensus decision never confirmed, so all new leader installation attempts fail, and the system stalls (a liveness violation). Alternatively, a more subtle safety violation is also possible because an incorrect proof from the past (with no decision ever having committed) could be wrongly verified as being correct under the next quorum system's rules.
\textbf{In our solution}, we address this problem by binding every proof to the quorum system which it has produced and then always verifying against the right quorum system. We exploit that each configuration change (membership reconfiguration or AWARE optimization) already installs a new view with incremented view number. Here, we maintain a view history that maps each consensus instance to the view that decided it. And the view includes its quorum system. Since any reconfiguration is itself applied after some instance $c$ and under some old view $v$, the new view $v_{next}$ governs consensus instances starting at $c+1$. A consensus decision is tagged with the ID of its governing view (e.g., $v$) and proof validation (both in leader change and state transfer) looks up the view for the corresponding consensus instance and applies the rules of $v$'s included quorum system.
For state transfer, a catching-up replica needs to learn of past configuration changes. This means the view history must be transferred (e.g., included as part of the replica state inside a checkpoint) to the fetching replica. Otherwise the fetching replica could be unable to validate the proof of the most recent state and would be unable to install the next leader.

\section{Evaluation}
\label{sec:eval}
The framework of Sec.~\ref{sec:design} lets us investigate a question that prior work left open:

  \textit{Under identical consensus protocol, identical optimizations, identical client behavior, and identical network, how much does the choice of quorum system actually matter for client-observed latency in wide-area BFT replication?}

 \smallskip  \noindent We address this question through five experiments.
First, we compare the four quorum systems at a canonical resilience threshold
(Sec.~\ref{sec:eval-baseline}) and then
study the interaction between quorum systems and
optimizations across
resilience settings (Sec.~\ref{sec:eval-opts}).
Afterwards, we explore the recovery behavior after a leader
failure (Sec.~\ref{sec:eval-faults}),
 investigate the resilience-performance trade-off that FaB Paxos makes in a WAN (Sec.~\ref{sec:eval-fab})
and measure how client geography affects the latency of collecting client quorums, as needed for read-only requests (Sec.~\ref{sec:eval-clients}).

\begin{figure}[t]
  \centering
  \scalebox{0.7}{\input{plots/heatmap_en}}
  \caption{Median round-trip latency (ms) between the 16 AWS regions.}
  \label{fig:heatmap}
\end{figure}
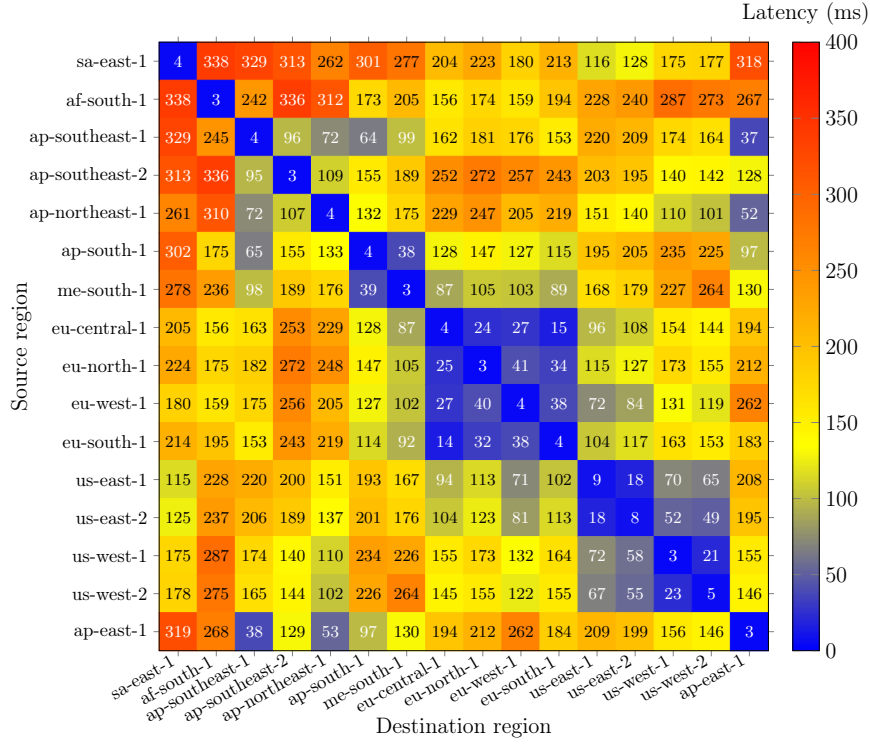

\subsection{Methodology}
\label{sec:eval-method}

To begin with, we briefly explain our methodology.

\para{Network environment and simulator.} We deploy $n=16$ replicas across 16 AWS regions covering
all continents (Fig.~\ref{fig:deployment}), Each also hosting one client to measure latency from different sites. As
the heatmap (Fig.~\ref{fig:heatmap}) shows, regions range from well-connected cliques (European regions have round-trip latencies below  50\,ms) to more isolated ones (e.g., South Africa) with intercontinental round-trips above 200\,ms. To simulate this WAN we use the Shadow network simulator~\cite{jansen2022co} which executes BFT-SMaRt replicas and clients as unmodified Linux processes and plugs them into a discrete-event simulation by interposing at the system call API, internally simulating TCP over a virtual network.
We prefer
Shadow over a real AWS deployment
because we previously found simulation results to come
close to a real deployment (\cite{berger2024chasingspeedlightlowlatency}, Fig.~14) and since
the simulation is deterministic, our results
are fully reproducible from the latency matrix in Fig.~\ref{fig:heatmap}
that
uses real latency statistics (the 50th percentile over one year).\footnote{As measured by Cloudping, see \url{https://www.cloudping.co/}}

\para{Workload, measurements and configurations.} Each client sends at least 1000 requests within 20 minutes of simulated time. When a client accepts a response, it waits for a randomized amount of time (0--200\,ms)\footnote{This is to ensure  clients in certain regions do not always win the race to have their requests included in the leader's next batch.}.
Clients measure their average end-to-end request latency, while replicas additionally measure consensus latency.
Unless stated otherwise, the client measurements are collected after the first round of self-optimization which AWARE applies every 500 consensus instances.
 In our first experiment, we choose $f=2$ as the canonical resilience setting because it is the largest value for $f$ at which a $\sqrt{n}\times\sqrt{n}$ grid is feasible for $n=16$. For later experiments (e.g., optimization and fault-injection), we additionally vary $f$ between 2 and 5 (threshold) or 2 and 4 (weighted, committee).

\begin{figure}[t]
  \centering
    \centering
    \input{plots/baseline_f2}
  \caption{Average client-observed latency per continent for $f=2$ \textit{after} optimization.
  }
  \label{fig:baseline-f2}
\end{figure}
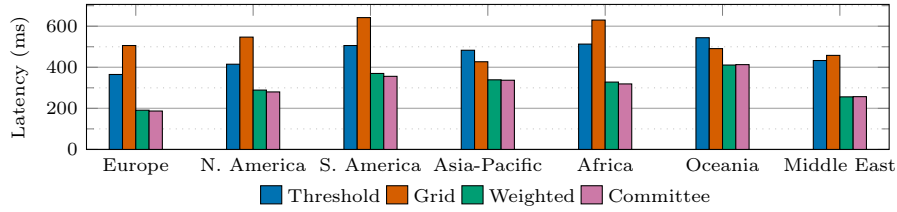
\subsection{Latency of Quorum System Constructions}
\label{sec:eval-baseline}

\para{Effect of the quorum system on client latency.} We begin with the head-to-head comparison of all four quorum system constructions at $f=2$ by measuring the average client latency grouped by continent (see Fig.~\ref{fig:baseline-f2}). In Fig.~\ref{fig:baseline-f2} we observe that across all regions, the weighted and committee quorums are faster than threshold quorums. This latency improvement is substantial in Europe (191/187\,ms for weighted/committee vs. 365\,ms for threshold) where clients are close to the replicas with high voting weight, but also visible (e.g., a 20--40\% reduction) for clients in other regions. Grid quorums perform slightly worse than threshold quorums.  The advantage of weighted and committee quorums comes from their minimal progress quorum of only  $2f+1=5$ replicas, which AWARE selects such that they are close to one another. Threshold quorums must wait for $\lceil(16+2+1)/2\rceil=10$ replicas including some intercontinental links.

\subsection{Combination with \emph{read-only} and \emph{tentative} Optimizations}
\label{sec:eval-opts}
\begin{figure}[t]
  \centering
  \input{plots/optimizations}
  \caption{Client-observed latency of \textit{ordered requests} averaged over all regions, for
  threshold (T), weighted (W), committee (C), and grid (G) quorums at
  varying $f$, under the three execution modes (normal/unoptimized,
  \emph{tentative}, \emph{read-only}).}
  \label{fig:opt}
\end{figure}
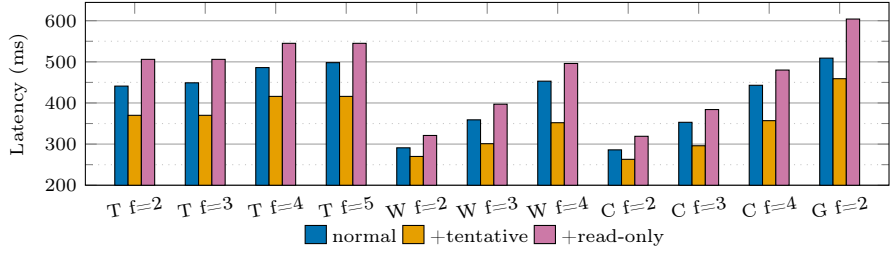

\para{Latency of write requests with optimizations.} In our next experiment, we study how our quorum systems interact with the \emph{read-only} and \emph{tentative} optimizations across $f\in\{2,3,4,5\}$. Fig.~\ref{fig:opt} displays the average client latency averaged over all regions. It is important to note that, in this experiment, we always measure the latency of ordered, i.e., \textit{write} requests. We measure read latencies in a later experiment.
Integrating the \emph{read-only} optimization thus always means observing a \textit{higher} latency, because clients are forced to wait for a quorum instead of just $f+1$ responses to their write requests. Integrating the \emph{tentative} optimization always leads to a lower latency because replicas reply to the client one step earlier, i.e., before the \textsc{Accept} stage.

\para{Observations.}
The \emph{tentative} requests optimization can accelerate all systems regardless of quorum system, but the effectiveness may be small. For instance, when quorums are collected fast as it is the case for weighted and committee quorums when $f=2$, then the effectiveness of the saved communication step is almost invisible (weighted from 291 to 270\,ms, committee from 286 to 263\,ms). This is because requiring a client to collect a quorum can offset some of the latency gains made when saving a communication step in the ordering path. As expected, integrating the \emph{read-only} optimization consistently degrades the performance of write requests (on average by 48\,ms or 11.7\%), because it imposes the client-quorum requirement on every write.  What is also clearly visible: Only the weighted quorum schemes, \textit{committee} and \textit{weighted}, are effective in trading resilience for improved latency.
At $f=2$, weighted and committee quorums are 34\% faster than threshold. At $f=4$, the gap is smaller but still clearly in their favor (453/443\,ms vs.\ 486\,ms in normal mode). The explanation comes back to the available $\Delta$: at $f=2$, weighted and committee have $\Delta=9$ spare replicas and allow a 5-replica progress quorum.  Threshold quorums can never be smaller than $\frac{n}{2}$ and thus are limited in how much they can trade resilience for latency.

\subsection{Behavior under Leader Failure}
\label{sec:eval-faults}

\para{Setup.} In this experiment, we crash the current leader at a fixed consensus instance and measure the client-observed latency between the crash and the first subsequent self-optimization round, averaged over the clients of each region. In this window the system still operates with a configuration tuned for the failed replica, which separates the cost of the failure from AWARE's ability to repair it. We compare weighted and committee quorum systems because both reach the same steady-state latency in fault-free operation (Sec.~\ref{sec:eval-baseline}).

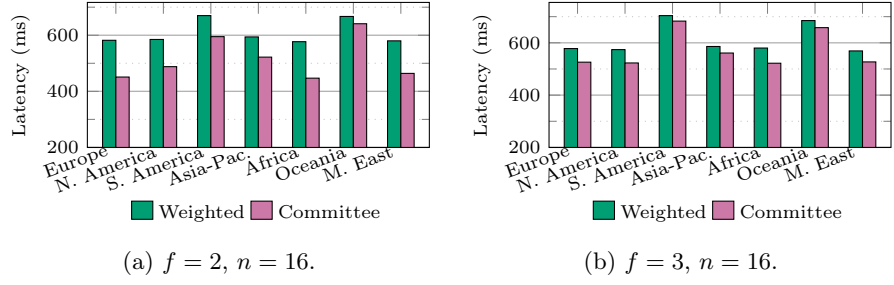
\begin{figure}[t]
  \centering
  \begin{subfigure}[t]{0.5\linewidth}
    \centering
    \input{plots/after_crashf2}
    \caption{$f=2$, $n=16$.}
    \label{fig:after-crash-f2}
  \end{subfigure}\hfill
  \begin{subfigure}[t]{0.5\linewidth}
    \centering
    \input{plots/after_crashf3}
    \caption{$f=3$, $n=16$.}
    \label{fig:after-crash-f3}
  \end{subfigure}
  \caption{Average client-observed latency in the post-failure,
  pre-optimization window after a leader crash. Committee quorums may collect a faster next-available
  quorum than weighted quorums. The gap narrows in most regions as $f$ grows.}
  \label{fig:faults}
\end{figure}

\para{Observations.} Committee quorums display a lower latency than weighted quorums in every region and at both resilience thresholds, but the difference decreases in most regions as $f$ grows (Fig.~\ref{fig:faults}). At $f=2$ we measure 451 vs.\ 582\,ms in Europe, 488 vs.\ 585 in North America, and 595 vs.\ 670 in South America. The constructions differ in the size of the next available quorum once a $V_{max}$ replica is crashed: In a weighted quorum its weight must be compensated by several $V_{min}$ replicas, which for $\Delta=9$ and $V_{max}=5.5$ grows the smallest quorum from $2f+1=5$ to $10$ replicas, whereas a committee quorum replaces the failed member by one of the remaining $3f+1$ nearby members and keeps the size at $5$. At $f=3$ the smaller slack ($\Delta=6$, $V_{max}=3$) grows that quorum only from $7$ to $9$ replicas, and the margin roughly halves (e.g., from 131 to 52\,ms in Europe). This latency penalty is transient: At the next optimization step, AWARE redistributes weights such that the weighted quorum system can catch up.

\subsection{The FaB Paxos Trade-off: Fewer Phases, Larger Quorums}
\label{sec:eval-fab}

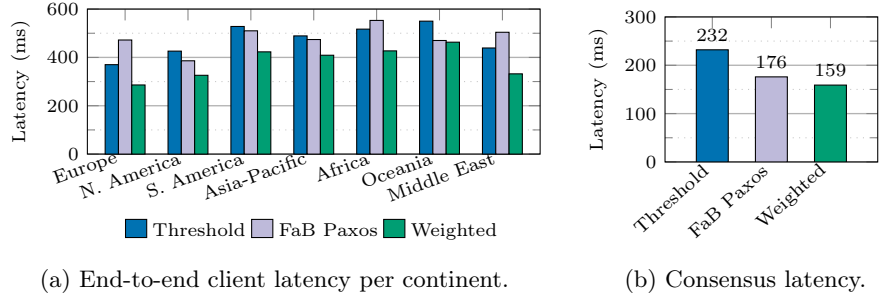
\begin{figure}[t]
  \centering
  \begin{subfigure}{0.62\linewidth}
    \centering
    \input{plots/fab}
    \caption{End-to-end client latency per continent.}
    \label{fig:fab-bars}
  \end{subfigure}\hfill
  \begin{subfigure}{0.36\linewidth}
    \centering
    \input{plots/fab_consensus}
    \caption{Consensus latency.}
    \label{fig:fab-consensus}
  \end{subfigure}
  \caption{The FaB Paxos message-pattern trade-off, at $f=3$, $n=16$.
  FaB Paxos is faster than threshold quorums in terms of pure
  consensus latency at the leader (b), but \emph{not} uniformly faster
  in terms of end-to-end client latency (a).}
  \label{fig:fab}
\end{figure}

We move to another question: \textit{Does saving a communication
step pay for proportionally larger quorums in a WAN?}
Fig.~\ref{fig:fab} compares threshold quorums, weighted quorums, and the FaB Paxos variant at $f=3$, the largest resilience setting at which FaB's optimal $n\geq 5f+1$ is achievable when $n=16$.
We also use $f=3$ for evaluating systems using the threshold and weighted quorums so that our comparison isolates the effect of the quorum size and message pattern.

\para{Observations.} At the leader, FaB Paxos does what it promises: The consensus latency drops from 232\,ms under threshold to 176\,ms under FaB, a 24\% reduction (Fig.~\ref{fig:fab-consensus}). Note that this is much lower than a naively expected 33\% reduction when saving one out of three communication steps. The reason lies in the proportionally larger quorums: $\lceil(16+10)/2\rceil = 13$ replicas at $f=3$ (FaB), compared to $\lceil(16+4)/2\rceil = 10$ for threshold. When the 13-replica quorum spans intercontinental links that the 10-replica quorum can avoid, replicas need to wait longer, which offsets part of the latency gain from the saved communication step.
Interestingly, systems with weighted quorums that still use the 3-step pattern but a smaller quorum do even better at 159\,ms.
When observing client latency (Fig.~\ref{fig:fab-bars}), FaB is faster than threshold in some regions and \emph{slower} in others.
 Weighted quorums win on average, by a 1.2\,$\times$--1.3\,$\times$ end-to-end speedup over threshold.
We conclude that the label \emph{fast} in Fast Byzantine Paxos applies mainly to \emph{consensus} latency at the leader (and perhaps to homogeneous-latency settings such as LANs)
 but does not necessarily speed up clients in WANs.
 When trading resilience for latency in a WAN,
 weighted replication in the normal 3-step protocol can beat a 2-step protocol like FaB.
 Junqueira et al.~\cite{classicvsfast} make a similar observation (fewer communication steps are not always better)
 for Classic vs.\ Fast Paxos in the crash-fault setting.

\subsection{Client Quorums and Geographic Effects}
\label{sec:eval-clients}

A recurring theme across all of our experiments was that consensus latency and client-perceived latency are not the same thing, and that the size (and structure) of the \emph{client-side} quorums matter as soon as \emph{read-only} or \emph{tentative} is enabled. To analyze that effect, we now inspect the client quorums in isolation.
We run a final experiment in which exclusively \emph{read-only} requests are issued, so that the consensus pattern is skipped and the client latency observed is just the time a client needs to collect its quorum.
\textbf{Setup.}
After a warm-up phase of sufficiently many ($\approx 300$ for each client) write requests to allow AWARE to optimize its configuration (quorum system and leader), each client issues read-only requests.  We use $f=3$ as the identical resilience setting for all quorum systems.

\begin{figure}[t]
  \centering
 \input{plots/reads}
  \caption{Consensus-free, \emph{read-only} client latencies per continent at $f=3$.
 FaB reads are the slowest from every region. Systems that use committee and weighted quorums
  win in \colorbox{green!20}{continents} where the client is closer to the cluster of $V_{max}$-replicas, but likewise can lose ground in the other \colorbox{red!20}{continents}.}
  \label{fig:clients}
\end{figure}
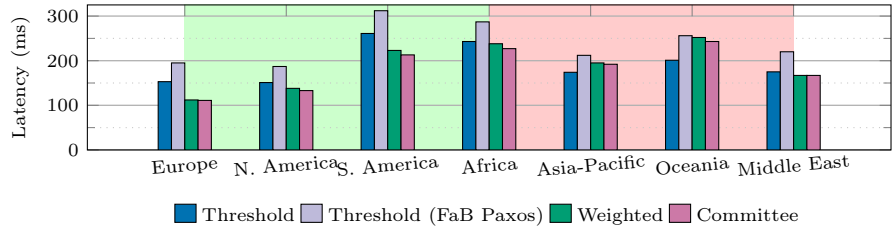

\para{Observations.}
FaB Paxos, with its 13-replica client quorum,
is the slowest in every region
(36--55\,ms behind threshold's 10-replica quorum). Weighted and committee quorums, which assign high
weight to European replicas, are collected fastest when clients are close to those replicas (in Europe, committee reads complete in
111\,ms; in North America, in 133\,ms). But in Asia-Pacific and Oceania, clients are far from high-weight replicas, so weighted and committee end up slightly slower than threshold (192 vs.\ 174\,ms in Asia-Pacific, 243 vs.\ 201 in Oceania).
Thus AWARE's consensus-optimal self-optimization, does not guarantee
read-optimal behavior for every client region. A system whose read workload is concentrated
far from the $V_{max}$-cluster may benefit from a different optimization target, or from
letting clients accept a result
under relaxed consistency without a full quorum (e.g., per-request) as in~\cite{guerraoui2016incremental}.

\section{Related Work}
\label{sec:relwork}

\para{BFT SMR frameworks and optimizations.}
BFT-SMaRt~\cite{bessani2014state} inherits several PBFT~\cite{castro1999practical} optimizations. The effectiveness of tentative executions in a WAN was studied by WHEAT~\cite{sousa2015separating} and the read-only optimization was revisited in~\cite{berger21reads} and later work used BFT-SMaRt as an ordering service for permissioned blockchains~\cite{sousa2018byzantine}.
These works improve the protocol or its message pattern; we instead
fix the optimizations and study how different quorum-system constructions affect WAN latency under otherwise identical conditions.

\para{Quorum Systems.} Malkhi and Reiter created a survey on Byzantine quorum systems~\cite{malkhi1997byzantine,malkhi1998byzantine} and
Bazzi showed synchrony assumptions permit smaller quorums~\cite{bazzi1997synchronous}. Probabilistic quorum systems relax intersection to a probabilistic guarantee~\cite{malkhi1997probabilistic},
recently made practical for BFT SMR to enable smaller quorums~\cite{avelas2024probabilistic} while
weighted quorums improve WAN latency by giving well-connected replicas higher voting weight~\cite{sousa2015separating}.
Asymmetric quorum systems let each replica specify its own fail-prone sets (i.e., trust assumptions) instead of using uniform assumptions~\cite{alpos2024asymmetricdistributedtrust}. In the crash-fault model, Flexible Paxos~\cite{howard2016flexible} relaxes intersection to hold only between leader-election and replication quorums, which WPaxos~\cite{ailijiang2019wpaxos} leverages for achieving improved latency in the WAN.
Our contribution is orthogonal: We quantify how selected quorum constructions compare in client-observed WAN latency inside a shared  BFT SMR implementation.

\para{Adaptive and reconfigurable BFT.} Many systems can adapt a SMR system to its environment during runtime, but might differ in what they adapt. ARCHER~\cite{eischer2018latency} selects the leader based on end-to-end latencies measured by clients, Fluidity~\cite{kostler2023fluidity} migrates replicas towards the geographic origin of incoming client workload, and OptiLog~\cite{gogada2026optilog} provides a logging framework for collecting and analyzing measurements for improved protocol role assignments in WANs despite faults.
BEWARE~\cite{chotkan2026robust} hardens AWARE-style weight reconfiguration against falsified latency reports from Byzantine replicas.
Carvalho et al. switch between entire consensus protocols at runtime~\cite{carvalho2018dynamic}.
Our framework instead makes the quorum system exchangeable and runtime-tunable, and shows how quorum-system transitions integrate correctly  with leader change and state transfer.

\para{Geo-replicated ordering.} Another line of research aims to improve wide-area replication.
Steward~\cite{amir10steward} proposes a hierarchical architecture, where several BFT system sites are connected over a  crash-fault tolerant (CFT) protocol in the WAN.
The CFT protocol Mencius~\cite{mao2008mencius} rotates the leader among sites where clients issue requests to the geographically closest replica.
Egalitarian Paxos~\cite{moraru2013there} has no dedicated leader (any replica can coordinate a request) and it orders only the requests that conflict.
 Coelho and Pedone build geo-replicated SMR on top of (crash-fault) atomic multicast~\cite{coelho2017fast,coelho18geopaxos}, which ByzCast transfers to the Byzantine fault model~\cite{coelho2018byzantine}.
 These systems propose an ordering protocol itself while we employ BFT-SMaRt's leader-based ordering pattern and vary only the quorum system, so that latency differences remain attributable to quorum design.

\section{Conclusion}
\label{sec:conclusion}
Quorum systems are a frequently overlooked degree of freedom in BFT SMR. In WAN deployments where the slowest replica of a quorum impacts the latency of collecting quorums during the ordering protocol, choosing the right quorum system (and its quorum formation rules) can make a real difference. Our BFT-SMaRt/AWARE variant makes the quorum system easy to swap or extend, and
our evaluation shows this choice directly impact client-observed latency.
Weighted and committee quorum systems can trade resilience for latency most effectively by leveraging proportionally smaller quorums. Threshold quorums cannot achieve that. Using protocols like FaB Paxos that shorten consensus by one communication step but need a proportionally larger threshold quorum in the WAN can actually lead to a higher latency than just using the normal 3-step protocol with weighted quorums at the same resilience threshold. Further, the consensus-optimal quorum is also not always client-optimal. Once read-only or tentative optimizations force clients to collect a full quorum, the client geography matters since clients far from a high-weight cluster might see no improvements.

\begin{credits}
\subsubsection{\ackname}
This work is funded by Deutsche Forschungsgemeinschaft (DFG, German Research Foundation) -- 446811880 (BFT2Chain).
\end{credits}

\bibliographystyle{splncs04}
\bibliography{references}

\end{document}

%% file: figures/normalbetrieb.tex
\begin{tikzpicture}[
  x=0.9cm,y=0.85cm,
  font=\small,
  >=latex,
  line cap=round,
  line join=round,
  msg/.style={->,thick,shorten <=1.5pt,shorten >=1.5pt},
  sep/.style={gray!70,densely dashed,thick},
  tline/.style={black,thick},
  stage/.style={font=\small\bfseries},
  quorum/.style={draw=blue!60!black,thick,dashed,
                 fill=blue!50,fill opacity=0.30,
                 rounded corners=4pt},
]
  \def\yC{4}
  \def\yRzero{3}
  \def\yRone{2}
  \def\yRtwo{1}
  \def\yRthree{0}

  \def\xA{0}
  \def\xB{2}
  \def\xC{4}
  \def\xD{6}
  \def\xE{8}
  \def\xF{10}

  \node[anchor=east] at (\xA-0.2,\yC) {C};
  \node[anchor=east] at (\xA-0.2,\yRzero) {R0};
  \node[anchor=east] at (\xA-0.2,\yRone) {R1};
  \node[anchor=east] at (\xA-0.2,\yRtwo) {R2};
  \node[anchor=east] at (\xA-0.2,\yRthree) {R3};

  \foreach \yy in {\yC,\yRzero,\yRone,\yRtwo,\yRthree} {
    \draw[tline] (\xA,\yy) -- (\xF+0.2,\yy);
  }

  \foreach \xx in {\xB,\xC,\xD,\xE,\xF} {
    \draw[sep] (\xx,\yRthree-0.25) -- (\xx,\yC+0.25);
  }

  \node[stage] at ({(\xA+\xB)/2},\yC+0.55) {Request};
  \node[stage] at ({(\xB+\xC)/2},\yC+0.55) {\textsc{Propose}};
  \node[stage] at ({(\xC+\xD)/2},\yC+0.55) {\textsc{Write}};
  \node[stage] at ({(\xD+\xE)/2},\yC+0.55) {\textsc{Accept}};
  \node[stage] at ({(\xE+\xF)/2},\yC+0.55) {Reply};

  \foreach \yy in {\yRzero, \yRone, \yRtwo, \yRthree} {
    \draw[msg] (\xA+0.6,\yC) -- (\xB,\yy);
  }

  \foreach \yy in {\yRone,\yRtwo,\yRthree} {
    \draw[msg] (\xB+0.05,\yRzero) -- (\xC,\yy);
  }

  \draw[msg] (\xC+0.15,\yRzero)   -- (\xD,\yRone);
  \draw[msg] (\xC+0.10,\yRzero)   -- (\xD,\yRtwo);
  \draw[msg] (\xC+0.05,\yRzero)   -- (\xD,\yRthree);

  \draw[msg] (\xC+0.15,\yRone)   -- (\xD,\yRzero);
  \draw[msg] (\xC+0.10,\yRone)   -- (\xD,\yRtwo);
  \draw[msg] (\xC+0.05,\yRone)   -- (\xD,\yRthree);

  \draw[msg] (\xC+0.05,\yRtwo)   -- (\xD,\yRzero);
  \draw[msg] (\xC+0.10,\yRtwo)   -- (\xD,\yRone);
  \draw[msg] (\xC+0.15,\yRtwo)   -- (\xD,\yRthree);

  \draw[msg] (\xC+0.05,\yRthree) -- (\xD,\yRzero);
  \draw[msg] (\xC+0.10,\yRthree) -- (\xD,\yRone);
  \draw[msg] (\xC+0.15,\yRthree) -- (\xD,\yRtwo);

  \foreach \yy in {\yRone,\yRtwo,\yRthree} { \draw[msg] (\xD+0.05,\yRzero) -- (\xE,\yy); }
  \draw[msg] (\xD+0.15,\yRone)   -- (\xE,\yRzero);
  \draw[msg] (\xD+0.10,\yRone)   -- (\xE,\yRtwo);
  \draw[msg] (\xD+0.05,\yRone)   -- (\xE,\yRthree);
  \draw[msg] (\xD+0.05,\yRtwo)   -- (\xE,\yRzero);
  \draw[msg] (\xD+0.10,\yRtwo)   -- (\xE,\yRone);
  \draw[msg] (\xD+0.15,\yRtwo)   -- (\xE,\yRthree);
  \draw[msg] (\xD+0.05,\yRthree) -- (\xE,\yRzero);
  \draw[msg] (\xD+0.10,\yRthree) -- (\xE,\yRone);
  \draw[msg] (\xD+0.15,\yRthree) -- (\xE,\yRtwo);

  \draw[msg] (\xE+0.05,\yRzero)  -- (\xF-1.10,\yC);
  \draw[msg] (\xE+0.05,\yRone)   -- (\xF-0.75,\yC);
  \draw[msg] (\xE+0.05,\yRtwo)   -- (\xF-0.40,\yC);
  \draw[msg] (\xE+0.05,\yRthree) -- (\xF+0.15,\yC);

  \draw[quorum] (\xC+0.08,\yRtwo-0.3) rectangle (\xD-0.08,\yRzero+0.3);
  \draw[quorum] (\xD+0.08,\yRtwo-0.3) rectangle (\xE-0.08,\yRzero+0.3);
  \draw[quorum] (\xF-1.35,\yC-0.28) rectangle (\xF-0.18,\yC+0.28);
\end{tikzpicture}

%% file: figures/threshold-quorum.tex
\begin{tikzpicture}[
  font=\small,
  line cap=round,
  line join=round,
  decorate=false,
  every path/.style={decorate=false},
  quorumG/.style={
    draw=green!70!black, very thick,
    dashed, dash pattern=on 12pt off 7pt,
    fill=green!55, fill opacity=0.45,
    rounded corners=2pt
  },
  quorumB/.style={
    draw=blue!60!black, very thick,
    dashed, dash pattern=on 12pt off 7pt,
    fill=blue!45, fill opacity=0.45,
    rounded corners=2pt
  }
]

\newcommand{\db}[5]{%
  \begin{scope}[shift={(#1,#2)}, scale=#3]

    \def\xr{1.35}
    \def\yr{0.38}
    \def\h{1.85}

    \path[fill=#4]
      (-\xr,0) -- (-\xr,-\h)
      arc[start angle=180, end angle=360, x radius=\xr, y radius=\yr]
      -- (\xr,0)
      arc[start angle=0, end angle=180, x radius=\xr, y radius=\yr]
      -- cycle;

    \draw[line width=2pt]
      (0,0) ellipse [x radius=\xr, y radius=\yr];

    \draw[line width=2pt] (-\xr,0) -- (-\xr,-\h);
    \draw[line width=2pt] ( \xr,0) -- ( \xr,-\h);

    \draw[line width=2pt]
      (0-\xr,-\h) arc[start angle=180, end angle=360,
                      x radius=\xr, y radius=\yr];

    \node[scale=#3 * 3] at (0,-1) {#5};

  \end{scope}%
}

\path[quorumG] (0,0) rectangle (12,7);
\path[quorumB] (4,-0.5) rectangle (16,6.5);

\db{2}{5.5}{0.8}{white}{}

\db{6}{5.5}{0.8}{white}{}

\db{10}{5.5}{0.8}{white}{}

\db{10}{2.5}{0.8}{white}{}

\db{14}{2.5}{0.8}{white}{}

\end{tikzpicture}

%% file: figures/weighted-quorum.tex
\begin{tikzpicture}[
  font=\small,
  line cap=round,
  line join=round,
  decorate=false,
  every path/.style={decorate=false},
  quorumG/.style={
    draw=green!70!black, very thick,
    dashed, dash pattern=on 12pt off 7pt,
    fill=green!55, fill opacity=0.45,
    rounded corners=2pt
  },
  quorumB/.style={
    draw=blue!60!black, very thick,
    dashed, dash pattern=on 12pt off 7pt,
    fill=blue!45, fill opacity=0.45,
    rounded corners=2pt
  }
]
\newcommand{\db}[5]{%
  \begin{scope}[shift={(#1,#2)}, scale=#3]
    \def\xr{1.35}
    \def\yr{0.38}
    \def\h{1.85}
    \path[fill=#4]
      (-\xr,0) -- (-\xr,-\h)
      arc[start angle=180, end angle=360, x radius=\xr, y radius=\yr]
      -- (\xr,0)
      arc[start angle=0, end angle=180, x radius=\xr, y radius=\yr]
      -- cycle;
    \draw[line width=2pt]
      (0,0) ellipse [x radius=\xr, y radius=\yr];
    \draw[line width=2pt] (-\xr,0) -- (-\xr,-\h);
    \draw[line width=2pt] ( \xr,0) -- ( \xr,-\h);
    \draw[line width=2pt]
      (0-\xr,-\h) arc[start angle=180, end angle=360,
                      x radius=\xr, y radius=\yr];
    \node[scale=#3 * 7, text=black] at (0,-1) {\contour{white}{#5}};
  \end{scope}%
}
\path[quorumG] (0,0) rectangle (12,7);
\path[quorumB] (8,-0.5) rectangle (16,6.5);
\db{2}{5.5}{0.8}{white}{1}
\db{6}{5.5}{0.8}{white}{1}
\db{10}{5.5}{0.8}{white}{1}
\db{10}{2.5}{0.8}{white}{2}
\db{14}{2.5}{0.8}{white}{2}
\end{tikzpicture}

%% file: figures/committee-quorum.tex
\begin{tikzpicture}[
  font=\small,
  line cap=round,
  line join=round,
  decorate=false,
  every path/.style={decorate=false},
  quorumG/.style={
    draw=green!70!black, very thick,
    dashed, dash pattern=on 12pt off 7pt,
    fill=green!55, fill opacity=0.45,
    rounded corners=2pt
  },
  quorumB/.style={
    draw=blue!60!black, very thick,
    dashed, dash pattern=on 12pt off 7pt,
    fill=blue!45, fill opacity=0.45,
    rounded corners=2pt
  },
  quorumW/.style={
    draw=black!60!black, very thick,
    dashed, dash pattern=on 12pt off 7pt,
    fill=black!45, fill opacity=0.45,
    rounded corners=2pt
  }
]
\newcommand{\db}[5]{%
  \begin{scope}[shift={(#1,#2)}, scale=#3]
    \def\xr{1.35}
    \def\yr{0.38}
    \def\h{1.85}
    \path[fill=#4]
      (-\xr,0) -- (-\xr,-\h)
      arc[start angle=180, end angle=360, x radius=\xr, y radius=\yr]
      -- (\xr,0)
      arc[start angle=0, end angle=180, x radius=\xr, y radius=\yr]
      -- cycle;
    \draw[line width=2pt]
      (0,0) ellipse [x radius=\xr, y radius=\yr];
    \draw[line width=2pt] (-\xr,0) -- (-\xr,-\h);
    \draw[line width=2pt] ( \xr,0) -- ( \xr,-\h);
    \draw[line width=2pt]
      (0-\xr,-\h) arc[start angle=180, end angle=360,
                      x radius=\xr, y radius=\yr];
    \node[scale=#3 * 7, text=black] at (0,-1) {\contour{white}{#5}};
  \end{scope}%
}
\path[quorumW] (0,0) rectangle (3.9,7);
\path[quorumG] (4.1,0) rectangle (12,7);
\path[quorumB] (8,-0.5) rectangle (16,6.5);
\db{2}{5.5}{0.8}{white}{0}
\db{6}{5.5}{0.8}{white}{1}
\db{10}{5.5}{0.8}{white}{1}
\db{10}{2.5}{0.8}{white}{1}
\db{14}{2.5}{0.8}{white}{1}
\end{tikzpicture}

%% file: figures/grid-quorums3x3.tex
\begin{tikzpicture}[
  font=\small,
  line cap=round,
  line join=round,
  decorate=false,
  every path/.style={decorate=false},
  quorumG/.style={
    draw=green!70!black, very thick,
    dashed, dash pattern=on 12pt off 7pt,
    fill=green!55, fill opacity=0.45,
    rounded corners=2pt
  },
  quorumB/.style={
    draw=blue!60!black, very thick,
    dashed, dash pattern=on 12pt off 7pt,
    fill=blue!45, fill opacity=0.45,
    rounded corners=2pt
  }
]
\newcommand{\db}[5]{%
  \begin{scope}[shift={(#1,#2)}, scale=#3]
    \def\xr{1.35}
    \def\yr{0.38}
    \def\h{1.85}
    \path[fill=#4]
      (-\xr,0) -- (-\xr,-\h)
      arc[start angle=180, end angle=360, x radius=\xr, y radius=\yr]
      -- (\xr,0)
      arc[start angle=0, end angle=180, x radius=\xr, y radius=\yr]
      -- cycle;
    \draw[line width=2pt]
      (0,0) ellipse [x radius=\xr, y radius=\yr];
    \draw[line width=2pt] (-\xr,0) -- (-\xr,-\h);
    \draw[line width=2pt] ( \xr,0) -- ( \xr,-\h);
    \draw[line width=2pt]
      (0-\xr,-\h) arc[start angle=180, end angle=360,
                      x radius=\xr, y radius=\yr];
    \node[scale=#3 * 3] at (0,-1) {#5};
  \end{scope}%
}
\path[quorumG] (-1.5,-1.25) rectangle (6.5,0.25);
\path[quorumG] (-0.75,-1.75) rectangle (0.75,4.75);
\path[quorumB] (-1.5,2.75) rectangle (6.5,4.25);
\path[quorumB] (1.75,-1.75) rectangle (3.25,4.75);
\db{0}{0}{0.5}{white}{}
\db{0}{2}{0.5}{white}{}
\db{0}{4}{0.5}{white}{}
\db{2.5}{0}{0.5}{white}{}
\db{2.5}{2}{0.5}{white}{}
\db{2.5}{4}{0.5}{white}{}
\db{5}{0}{0.5}{white}{}
\db{5}{2}{0.5}{white}{}
\db{5}{4}{0.5}{white}{}
\end{tikzpicture}

%% file: tables/table1.tex
\begin{tabular}{lccccc}
    \toprule
    \textbf{Quorum System} &
    \textbf{Resilience} &
    \textbf{\# Steps} &
    \textbf{Quorum} &
    \textbf{Client} &
    \textbf{Tentative} \\
    \midrule

    \multicolumn{6}{l}{\textbf{Threshold Quorums}} \\
    \midrule
    Threshold
      &
      & 3
      &
      & $f+1$ replies
      & \\

    Threshold + ro
      & $3f+1 \leq n$
      & 3
      & $\lceil \frac{n+f+1}{2} \rceil$
      & Quorum
      & \\

    Threshold + ro + t
      &
      & 2
      &
      & Quorum
      & $\checkmark$ \\

    \midrule

    Fast Byzantine Paxos (FaB)
      & $5f+1 \leq n$
      & 2
      & $\lceil \frac{n+3f+1}{2} \rceil$
      & Quorum
      & \\

    \midrule
    \multicolumn{6}{l}{\textbf{Weighted Quorums (WHEAT)}} \\
    \midrule
    Weighted
      &
      & 3
      & \multirow{3}{*}{%
          \makecell[c]{$2f\cdot V_{max}+1$\\voting weight}}
      & $f+1$ replies
      & \\

    Weighted + ro
      & $3f + 1 + \Delta = n$
      & 3
      &
      & Quorum
      & \\

    Weighted + ro + t
      &
      & 2
      &
      & Quorum
      & $\checkmark$ \\

    \midrule
    \multicolumn{6}{l}{\textbf{Grid Quorums}} \\
    \midrule
    Square grid ($\sqrt{n} \times \sqrt{n}$)
      &
      & 3
      & \multirow{3}{*}{%
          \makecell[c]{$1$ col. $+\ \lceil \frac{f+1}{2} \rceil$ rows \\ $\lor~1$ row $+\ \lceil \frac{f+1}{2} \rceil$ col.}}
      & $f+1$ replies
      & \\

    Square grid ($\sqrt{n} \times \sqrt{n}$) + ro
      & $\lceil \frac{3f+1}{2} \rceil \leq \sqrt{n}$
      & 3
      &
      & Quorum
      & \\

    Square grid ($\sqrt{n} \times \sqrt{n}$) + ro + t
      &
      & 2
      &
      & Quorum
      & $\checkmark$ \\

    \midrule
    \multicolumn{6}{l}{\textbf{Committee Quorums}} \\
    \midrule
    Committee
      &
      & 3
      &
      & $f+1$ replies
      & \\

    Committee + ro
      & $3f + 1 + \Delta = n$
      & 3
      & $2f+1$ members
      & Quorum
      & \\

    Committee + ro + t
      &
      & 2
      &
      & Quorum
      & $\checkmark$ \\
    \bottomrule
\end{tabular}

%% file: plots/heatmap_en.tex
\centering
\begin{tikzpicture}
\begin{axis}[
    width=11.5cm,
    height=11.5cm,
    scale only axis,
    axis equal image,
    enlargelimits=false,
    xlabel={Destination region},
    ylabel={Source region},
    xlabel style={
        at={(axis description cs:0.5,-0.1)},
        anchor=north,
         font=\large,
    },
    ylabel style={
        at={(axis description cs:-0.20,0.5)},
        anchor=south,
         font=\large,
    },
    colorbar,
    colorbar style={
        title={Latency (ms)},
         font=\large,
        at={(1.04,0)},
        anchor=south west,
        height=\pgfkeysvalueof{/pgfplots/parent axis height}
    },
    point meta min=0,
    point meta max=400,
    xmin=-0.5, xmax=15.5,
    ymin=-0.5, ymax=15.5,
    xtick={0,...,15},
    ytick={0,...,15},
    xticklabels from table={data/aws-regions.dat}{[index]1},
    yticklabels from table={data/aws-regions.dat}{[index]1},
    tick align=outside,
    tick style={major tick length=3pt, thin},
    xticklabel style={
        font=\normalsize,
        rotate=30,
        anchor=east,
    },
    yticklabel style={font=\normalsize},
    clip=false,
]

\addplot[
    matrix plot,
    mesh/cols=16,
    point meta=explicit,
]
table[
    x index=0,
    y index=1,
    meta index=2,
    col sep=space,
] {data/aws-latency-values.dat};

\addplot[
    scatter,
    only marks,
    mark size=0pt,
    point meta=explicit,
    nodes near coords={\pgfmathprintnumber[fixed,precision=0]{\pgfplotspointmeta}},
    every node near coord/.style={
        font=\footnotesize,
        text=white,
        anchor=center,
    },
]
table[
    x index=0,
    y index=1,
    meta index=2,
    col sep=space,
] {data/aws-latency-values-white.dat};

\addplot[
    scatter,
    only marks,
    mark size=0pt,
    point meta=explicit,
    nodes near coords={\pgfmathprintnumber[fixed,precision=0]{\pgfplotspointmeta}},
    every node near coord/.style={
        font=\footnotesize,
        text=black,
        anchor=center,
    },
]
table[
    x index=0,
    y index=1,
    meta index=2,
    col sep=space,
] {data/aws-latency-values-black.dat};

\end{axis}
\end{tikzpicture}

%% file: plots/baseline_f2.tex
\begin{tikzpicture}
\begin{axis}[
    ybar,
    width=1\linewidth,
    height=3.5cm,
     tick align=inside,
    ylabel={Latency (ms)},
    symbolic x coords={
        Europe, N.~America, S.~America, Asia-Pacific,
        Africa, Oceania, Middle~East
    },
    ymin=0,
    xtick=data,
    xticklabel style={font=\scriptsize, rotate=0},
    yticklabel style={font=\scriptsize},
    ylabel style={font=\scriptsize},
    bar width=5pt,
    enlarge x limits=0.07,
    ymajorgrids=true,
    yminorgrids=true,
    minor y tick num=1,
    major grid style={solid, gray!60, line width=0.5pt},
    minor grid style={dotted, gray!45, line width=0.4pt},
    every axis plot/.append style={draw=black, line width=0.4pt},
    legend style={at={(0.5,-0.2)}, anchor=north,
                  legend columns=4, font=\scriptsize, draw=none},
    legend image code/.code={%
        \draw[#1, draw=black]
            (0cm,-0.08cm) rectangle (0.28cm,0.16cm);}
]
\addplot[fill=schwellwert, bar shift=-7.5pt]
    coordinates {(Europe,365)(N.~America,415)(S.~America,506)(Asia-Pacific,483)
                 (Africa,513)(Oceania,544)(Middle~East,433)};
\addplot[fill=gitter, bar shift=-2.5pt]
    coordinates {(Africa,630)(Middle~East,458)(Oceania,491)(S.~America,642)(N.~America,547)(Europe,506)(Asia-Pacific,427)};
\addplot[fill=gewichtet, bar shift=2.5pt]
    coordinates {(Africa,328)(Middle~East,256)(Oceania,411)(S.~America,370)(N.~America,289)(Europe,191)(Asia-Pacific,339)};
\addplot[fill=komitee, bar shift=7.5pt]
    coordinates {(Africa,319)(Middle~East,257)(Oceania,413)(S.~America,356)(N.~America,280)(Europe,187)(Asia-Pacific,337)};
\legend{Threshold, Grid, Weighted, Committee}
\end{axis}
\end{tikzpicture}
\vspace{-0.1cm}

%% file: plots/optimizations.tex
\begin{tikzpicture}
\begin{axis}[
    ybar,
    width=1\linewidth,
    height=4cm,
     tick align=inside,
    ylabel={Latency (ms)},
    symbolic x coords={
        T~f=2, T~f=3, T~f=4, T~f=5,
        W~f=2, W~f=3, W~f=4,
        C~f=2, C~f=3, C~f=4,
        G~f=2
    },
    ymin=200,
    xtick=data,
    xticklabel style={font=\scriptsize, rotate=10},
    yticklabel style={font=\scriptsize},
    ylabel style={font=\scriptsize},
    bar width=5pt,
    enlarge x limits=0.07,
    ymajorgrids=true,
    yminorgrids=true,
    minor y tick num=1,
    major grid style={solid, gray!60, line width=0.5pt},
    minor grid style={dotted, gray!45, line width=0.4pt},
    every axis plot/.append style={draw=black, line width=0.4pt},
    legend style={at={(0.5,-0.18)},  anchor=north,
                  legend columns=3, font=\scriptsize, draw=none},
    legend image code/.code={%
        \draw[#1, draw=black]
            (0cm,-0.08cm) rectangle (0.28cm,0.16cm);}
]
\addplot[fill=cnormal, bar shift=-5pt]
    coordinates {(T~f=2,441)(T~f=3,449)(T~f=4,486)(T~f=5,498)
                 (G~f=2, 509)
                 (C~f=2, 286) (C~f=3, 353) (C~f=4, 443)
                 (W~f=2, 291) (W~f=3, 359) (W~f=4, 453)};
\addplot[fill=ctentative, bar shift=0pt]
    coordinates {(T~f=2,370)(T~f=3,370)(T~f=4,416)(T~f=5,416)
                 (G~f=2, 459)
                 (C~f=2, 263) (C~f=3, 296) (C~f=4, 357)
                 (W~f=2, 270) (W~f=3, 301) (W~f=4, 352)};
\addplot[fill=creadonly, bar shift=5pt]
    coordinates {(T~f=2,506)(T~f=3,506)(T~f=4,545)(T~f=5,545)
                 (G~f=2, 604)
                 (C~f=2, 319) (C~f=3, 384) (C~f=4, 480)
                 (W~f=2, 321) (W~f=3, 397) (W~f=4, 496)};
\legend{normal, +tentative, +read-only}
\end{axis}
\end{tikzpicture}

%% file: plots/after_crashf2.tex
\begin{tikzpicture}
\begin{axis}[
    ybar,
    width=1\linewidth,
    height=3.5cm,
     tick align=inside,
    ylabel={Latency (ms)},
    symbolic x coords={
        Europe, N.~America, S.~America, Asia-Pac.,
        Africa, Oceania, M.~East
    },
    ymin=200,
    xtick=data,
    xticklabel style={font=\scriptsize, rotate=20, anchor=east},
    yticklabel style={font=\scriptsize},
    ylabel style={font=\scriptsize},
    bar width=5pt,
    enlarge x limits=0.10,
    ymajorgrids=true,
    yminorgrids=true,
    minor y tick num=1,
    major grid style={solid, gray!60, line width=0.5pt},
    minor grid style={dotted, gray!45, line width=0.4pt},
    every axis plot/.append style={draw=black, line width=0.4pt},
    legend style={at={(0.5,-0.32)}, anchor=north,
                  legend columns=2, font=\scriptsize, draw=none},
    legend image code/.code={%
        \draw[#1, draw=black]
            (0cm,-0.08cm) rectangle (0.28cm,0.16cm);}
]
\addplot[fill=gewichtet, bar shift=-2.5pt]
    coordinates {(Europe,582)(N.~America,585)(S.~America,670)(Asia-Pac.,594)
                 (Africa,577)(Oceania,667)(M.~East,580)};
\addplot[fill=komitee, bar shift=2.5pt]
    coordinates {(Europe,451)(N.~America,488)(S.~America,595)(Asia-Pac.,522)
                 (Africa,447)(Oceania,641)(M.~East,464)};
\legend{Weighted, Committee}
\end{axis}
\end{tikzpicture}

%% file: plots/after_crashf3.tex
\begin{tikzpicture}
\begin{axis}[
    ybar,
    width=1\linewidth,
    height=3.5cm,
     tick align=inside,
    ylabel={Latency (ms)},
    symbolic x coords={
        Europe, N.~America, S.~America, Asia-Pac.,
        Africa, Oceania, M.~East
    },
    ymin=200,
    xtick=data,
    xticklabel style={font=\scriptsize, rotate=20, anchor=east},
    yticklabel style={font=\scriptsize},
    ylabel style={font=\scriptsize},
    bar width=5pt,
    enlarge x limits=0.10,
    ymajorgrids=true,
    yminorgrids=true,
    minor y tick num=1,
    major grid style={solid, gray!60, line width=0.5pt},
    minor grid style={dotted, gray!45, line width=0.4pt},
    every axis plot/.append style={draw=black, line width=0.4pt},
    legend style={at={(0.5,-0.32)}, anchor=north,
                  legend columns=2, font=\scriptsize, draw=none},
    legend image code/.code={%
        \draw[#1, draw=black]
            (0cm,-0.08cm) rectangle (0.28cm,0.16cm);}
]
\addplot[fill=gewichtet, bar shift=-2.5pt]
    coordinates {(Europe,578)(N.~America,574)(S.~America,704)(Asia-Pac.,586)
                 (Africa,580)(Oceania,685)(M.~East,569)};
\addplot[fill=komitee, bar shift=2.5pt]
    coordinates {(Europe,526)(N.~America,523)(S.~America,683)(Asia-Pac.,561)
                 (Africa,522)(Oceania,658)(M.~East,527)};
\legend{Weighted, Committee}
\end{axis}
\end{tikzpicture}

%% file: plots/fab.tex
\begin{tikzpicture}
\begin{axis}[
    ybar,
    width=\linewidth,
    height=3.5cm,
    tick align=inside,
    ylabel={Latency (ms)},
    symbolic x coords={
        Europe, N.~America, S.~America, Asia-Pacific,
        Africa, Oceania, Middle~East
    },
    ymin=0,
    xtick=data,
    xticklabel style={font=\scriptsize, rotate=20, anchor=east},
    yticklabel style={font=\scriptsize},
    ylabel style={font=\scriptsize},
    bar width=5pt,
    enlarge x limits=0.1,
    ymajorgrids=true,
    yminorgrids=true,
    minor y tick num=1,
    ytick={0, 200,400,600},
    ymax=600,
    major grid style={solid, gray!60, line width=0.5pt},
    minor grid style={dotted, gray!45, line width=0.4pt},
    every axis plot/.append style={draw=black, line width=0.4pt},
    legend style={at={(0.5,-0.4)}, anchor=north,
                  legend columns=3, font=\scriptsize, draw=none},
    legend image code/.code={%
        \draw[#1, draw=black]
            (0cm,-0.08cm) rectangle (0.28cm,0.16cm);}
]
\addplot[fill=schwellwert, bar shift=-5pt]
    coordinates {(Europe,370)(N.~America,426)(S.~America,528)(Asia-Pacific,489)
                 (Africa,517)(Oceania,550)(Middle~East,439)};
\addplot[fill=fab, bar shift=0pt]
    coordinates {(Europe,472)(N.~America,386)(S.~America,510)(Asia-Pacific,474)
                 (Africa,553)(Oceania,470)(Middle~East,504)};
\addplot[fill=gewichtet, bar shift=5pt]
    coordinates {(Africa,427)(Middle~East,332)(Oceania,463)(S.~America,423)(N.~America,326)(Europe,286)(Asia-Pacific,409)}
;
\legend{Threshold, FaB Paxos, Weighted}
\end{axis}
\end{tikzpicture}

%% file: plots/fab_consensus.tex
\begin{tikzpicture}
\begin{axis}[
    ybar,
    bar width=12pt,
     tick align=inside,
    width=\linewidth,
    height=3.5cm,
    symbolic x coords={Threshold,FabPaxos,Weighted},
    xtick={Threshold,FabPaxos,Weighted},
    xticklabels={Threshold,{FaB Paxos},Weighted},
    ymin=0,
    ymax=300,
    ylabel={Latency (ms)},
    ylabel style={font=\scriptsize},
    xticklabel style={font=\scriptsize, rotate=42, anchor=east},
    yticklabel style={font=\scriptsize},
    enlarge x limits=0.4,
    nodes near coords,
    every node near coord/.append style={font=\scriptsize},
    ymajorgrids=true,
    yminorgrids=true,
    minor y tick num=1,
    major grid style={solid, gray!60, line width=0.5pt},
    minor grid style={dotted, gray!45, line width=0.4pt},
]
\addplot[bar shift=0pt, draw=black, line width=0.4pt, fill=schwellwert]
    coordinates {(Threshold,232)};
\addplot[bar shift=0pt, draw=black, line width=0.4pt, fill=fab]
    coordinates {(FabPaxos,176)};
\addplot[bar shift=0pt, draw=black, line width=0.4pt, fill=gewichtet]
    coordinates {(Weighted,159)};
\end{axis}
\end{tikzpicture}
\hfill

%% file: plots/reads.tex
\begin{tikzpicture}
\begin{axis}[
    ybar,
    width=\linewidth,
    height=3.5cm,
    tick align=inside,
    ylabel={Latency (ms)},
    symbolic x coords={
        Europe, N.~America, S.~America, Africa,  Asia-Pacific,
        Oceania, Middle~East
    },
    ymin=0,
    ymax=325,
    xtick=data,
    xticklabel style={font=\scriptsize, rotate=5},
    yticklabel style={font=\scriptsize},
    ylabel style={font=\scriptsize},
    bar width=5pt,
    enlarge x limits=0.16,
    ymajorgrids=true,
    yminorgrids=true,
    minor y tick num=1,
    major grid style={solid, gray!60, line width=0.5pt},
    minor grid style={dotted, gray!45, line width=0.4pt},
    every axis plot/.append style={draw=black, line width=0.4pt},
    legend style={at={(0.5,-0.32)}, anchor=north,
                  legend columns=4, font=\scriptsize, draw=none},
    legend image code/.code={%
        \draw[#1, draw=black]
            (0cm,-0.08cm) rectangle (0.28cm,0.16cm);}
]
\begin{scope}[on background layer]
    \fill[green!20] (axis cs:Europe,0) rectangle (axis cs:Africa,325);
    \fill[red!20]   (axis cs:Africa,0) rectangle (axis cs:Middle~East,325);
\end{scope}
\addplot[fill=schwellwert, bar shift=-7.5pt]
    coordinates {(Europe,153)(N.~America,151)(S.~America,261)(Asia-Pacific,174)
                 (Africa,243)(Oceania,201)(Middle~East,175)};
\addplot[fill=fab, bar shift=-2.5pt]
    coordinates {(Europe,195)(N.~America,187)(S.~America,312)(Asia-Pacific,212)
                 (Africa,287)(Oceania,256)(Middle~East,220)};
\addplot[fill=gewichtet, bar shift=2.5pt]
    coordinates {(Africa,238)(Middle~East,167)(Oceania,252)(S.~America,223)(N.~America,138)(Europe,112)(Asia-Pacific,195)};
\addplot[fill=komitee, bar shift=7.5pt]
    coordinates {(Africa,227)(Middle~East,167)(Oceania,243)(S.~America,213)(N.~America,133)(Europe,111)(Asia-Pacific,192)};
\legend{Threshold, Threshold (FaB Paxos), Weighted, Committee}
\end{axis}
\end{tikzpicture}